\documentclass{ws-p8-50x6-00}

\begin{document}

\title{Substructure of jets at HERA}

\author{M\'ONICA V\'AZQUEZ (on behalf of the ZEUS Collaboration)}
\address{Dpto. de F\'{\i}sica Te\'orica, Universidad Aut\'onoma de Madrid, Cantoblanco, E-28049
Madrid, Spain\\  
E-mail: monicava@mail.desy.de}


\maketitle
\abstracts{The substructure of jets produced in an exclusive and a charm-induced
 dijet sample in photoproduction and in charged and neutral current 
interactions has been studied with the ZEUS detector at HERA. Jets were 
identified using the longitudinally invariant $k_T$ cluster algorithm in the laboratory frame. 
The substructure of jets has been studied in terms of the jet shape and 
subjet multiplicity. Comparisons between the dijet sample and the 
quark-induced samples allow an extraction of the jet substructure for gluons. Leading-logarithm parton-shower Monte Carlo calculations give a 
good description of the differences between quark- and gluon-initiated jets. 
In neutral current interactions, the measurements have been compared 
to next-to-leading-order QCD calculations which are used to make a
determination of the strong coupling constant, $\alpha_s$. 
\vspace*{-0.8cm}
\newline
}

\section{Introduction}
\vspace*{-0.3cm}
The internal structure of jets gives insight into the transition between 
partons produced in the hard scattering process and the experimentally 
observable jets of hadrons. The substructure of jets is expected to depend
 mainly on the type of primary parton, quark or gluon, and to a lesser extent 
on the particular hard scattering process. QCD predicts that at high 
transverse energy of the jet, $E_T^{jet}$, where fragmentation effects 
become negligible, the jet structure is driven by gluon emission from the 
primary parton and gluon jets are broader than quark jets due to the larger 
colour charge of the gluon. The substructure of jets has been studied in terms 
of the integrated jet shape~\cite{sh}, $\Psi(r)$, and the differential jet shape~\cite{sh}, $\rho(r)$, where the energy flow inside a jet is considered, and the
subjet multiplicity~\cite{su}, $\langle n_{sbj}\rangle$, where jet-like structures (subjets) within a given jet are studied. The data samples used in these analyses were collected with the ZEUS detector at HERA. During 1995-1997 (1999-2000) HERA operated with positrons of energy $E_e=27.5$ GeV colliding with protons of energy $E_p=820$ GeV ($E_p=920$ GeV).

\begin{figure} [t]
\vspace*{-2.cm}
  \unitlength 1cm
\begin{minipage}{6cm}
  \begin{picture}(5,5)
    \put(0.,0.){\psfig{file=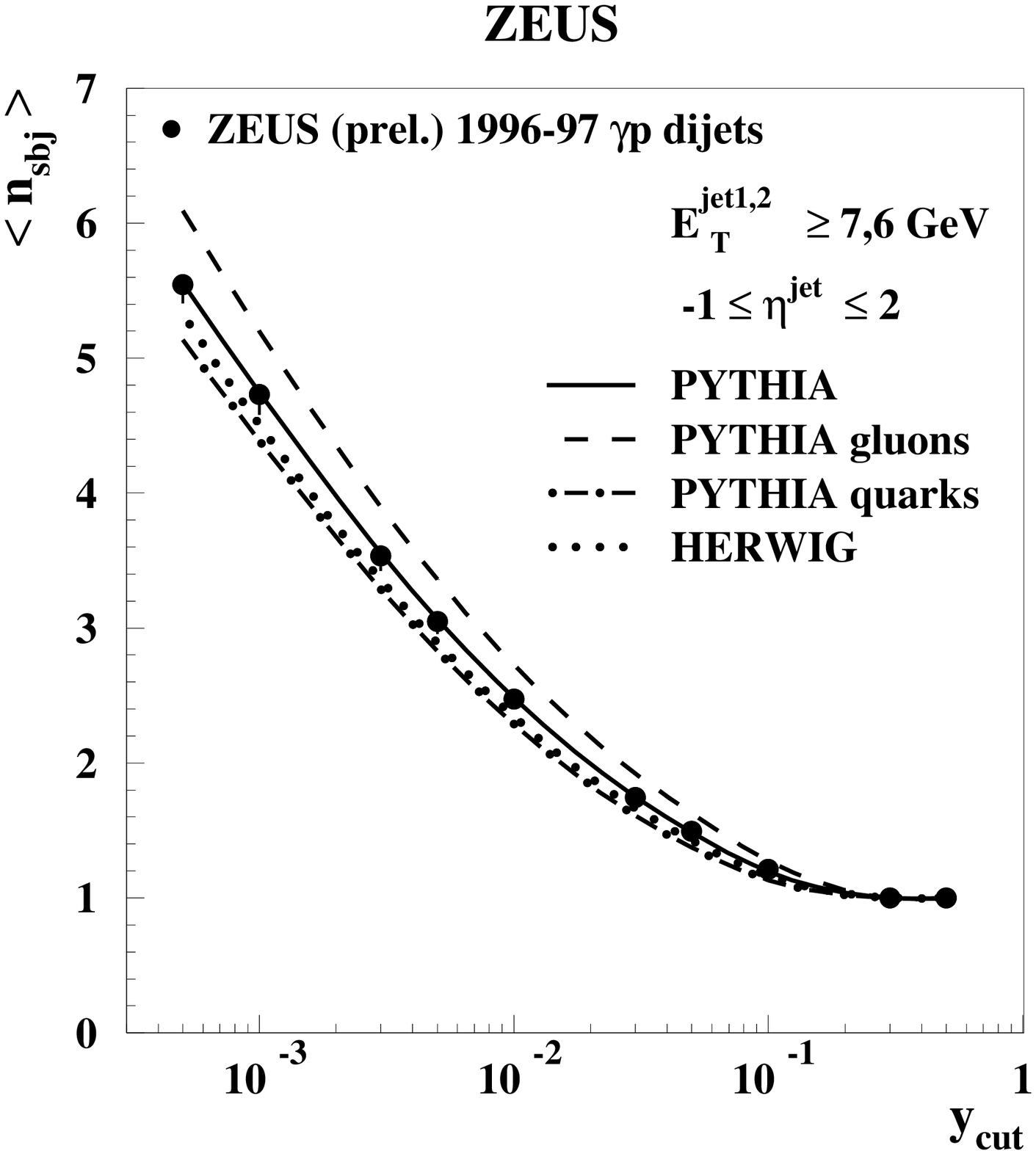,height=6.0cm,clip=}}
    \put(1.,1.){(a)}
  \end{picture}
\end{minipage}
\hfill
\begin{minipage}{6cm}
\begin{picture}(5,5)
    \put(0.,0.){\psfig{file=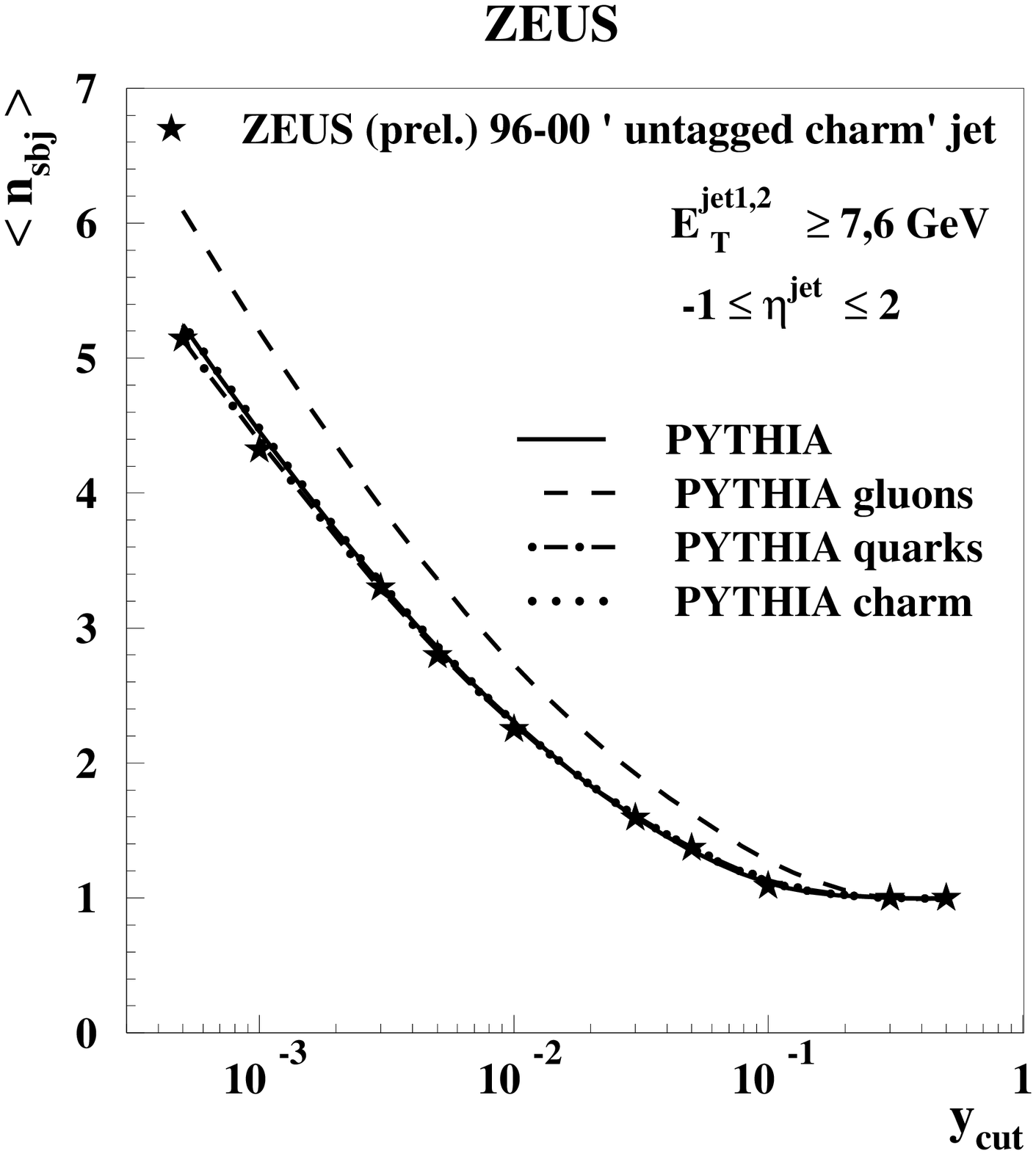,height=6.0cm,clip=}}
    \put(1.,1.){(b)}
  \end{picture}
\end{minipage}
\vspace*{-0.2cm}
\caption{
Measured $\langle n_{sbj}\rangle$  as a function of $y_{cut}$ corrected to hadron level for  (a) the exclusive dijet sample,  and (b) the ``untagged-charm'' jet in photoproduction.
\label{sub_ycut}
\vspace*{-0.8cm}
}
\end{figure}
\begin{figure}[b]
\vspace*{0.2cm}
  \unitlength 1cm
\begin{minipage}{6cm}
  \begin{picture}(5,5)
    \put(0.,0.){\psfig{file=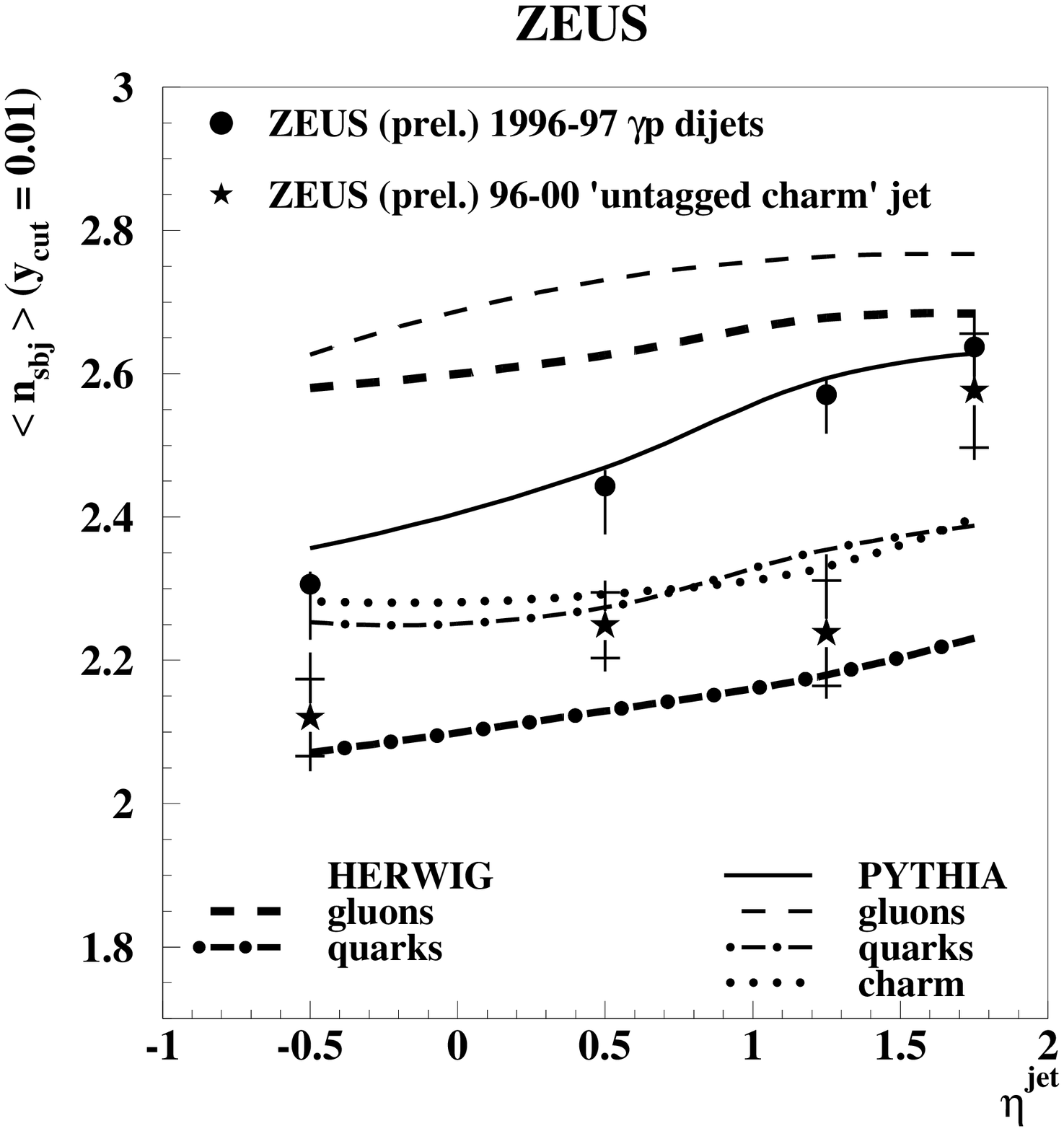,height=6.0cm,clip=}}
    \put(2.5,1.){(a)}
  \end{picture}
\end{minipage}
\hfill
\begin{minipage}{6cm}
  \begin{picture}(5,5)
    \put(0.,0.){\psfig{file=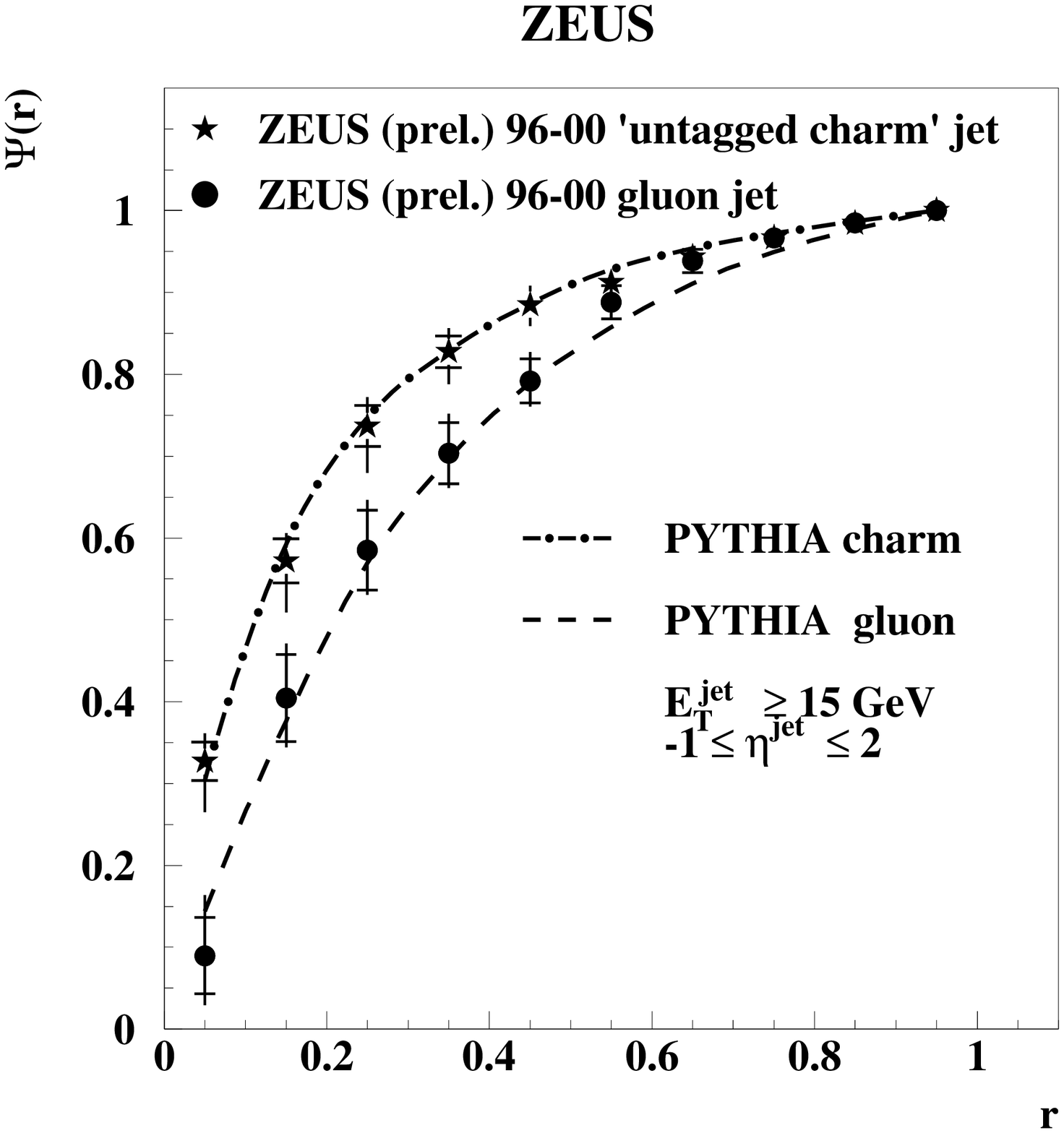,height=6.0cm,clip=}}
    \put(2.,1.){(b)}
  \end{picture}
\end{minipage}
\vspace*{-0.2cm}
\caption{
(a) $\langle n_{sbj}\rangle$ at a fixed $y_{cut}=0.01$ as a function of $\eta^{jet}$ corrected to hadron level for the exclusive dijet sample and the ``untagged-charm'' jet . (b) Integrated jet shape  of the measured ``untagged-charm'' jet and the extracted gluon jet substructure.
\label{sub_eta}
\vspace*{-.8cm}
}
\end{figure}
\vspace*{-0.3cm}
\section{Jet substructure in exclusive and charm-induced dijets in $\gamma p$}
\vspace*{-0.3cm}
Jet shapes and subjet multiplicities in exclusive and  charm-induced dijet samples in photoproduction (PHP) for the kinematic region $Q^2\leq 1$ GeV$^2$ and inelasticity $0.2<y<0.85$ have been measured. Dijets events were identified with the $k_T$-cluster 
algorithm~\cite{kt} and selected with $E_T^{jet_{1,2}}\geq 7,6$ GeV and $-1<\eta^{jet}<2$. Charm quarks were tagged by identifying $D^{*\pm}$ mesons through the $K2\pi$ decay mode
using the $\Delta M$ 
method
. The jet with closest distance in azimuthal angle to the $D^{*}$ meson was associated with the charm meson. The jet not associated with the $D^{*}$, referred to as  the ``untagged-charm'' jet, represents the unbiased, i.e. not influenced 
by the $D^{*}$ selection, jet candidate whose internal properties are studied.

The measured $\langle n_{sbj}\rangle$ as a function of the resolution scale, $y_{cut}$, for the exclusive dijet sample is shown in Fig.~\ref{sub_ycut}a. The predicted $\langle n_{sbj}\rangle$ is larger for gluon-initiated jets than for quark-initiated jets and the measured data are located between the two curves, showing that the dijet sample is a mixture of quark and gluon jets. PYTHIA~\cite{pythia} gives a good description of the data. Fig.~\ref{sub_ycut}b shows $\langle n_{sbj}\rangle$  as a function of $y_{cut}$ for the ``untagged-charm'' jet. The agreement between data and theory is very good and the predictions of charm-initiated jets are consistent with the measurements. 

The dependence of the substructure of jets with $\eta^{jet}$ has also been studied. In the exclusive dijet sample, $\langle n_{sbj}\rangle$ increases with $\eta^{jet}$ (Fig.~\ref{sub_eta}a), which is consistent with the predicted increase in the fraction of gluon-initiated jets with $\eta^{jet}$. In the ``untagged-charm'' jet sample,  the results are consistent with a pure sample of quark jets for $-1<\eta^{jet}<1.5$. For the highest $\eta^{jet}$ values, the data show 
a deviation from the prediction for quark-induced jets. Since the estimated gluon contamination to the charm-induced jet sample (mainly due to ``charm excitation'') has its highest contribution in the forward region, the deviation could be explained by the increase of the gluon-jet fraction in the charm-enriched sample.
\begin{figure}[t]
  \unitlength 1cm
\vspace*{-1.5cm}
\begin{minipage}{6cm}
  \begin{picture}(5,5)
    \put(0.,0.){\psfig{file=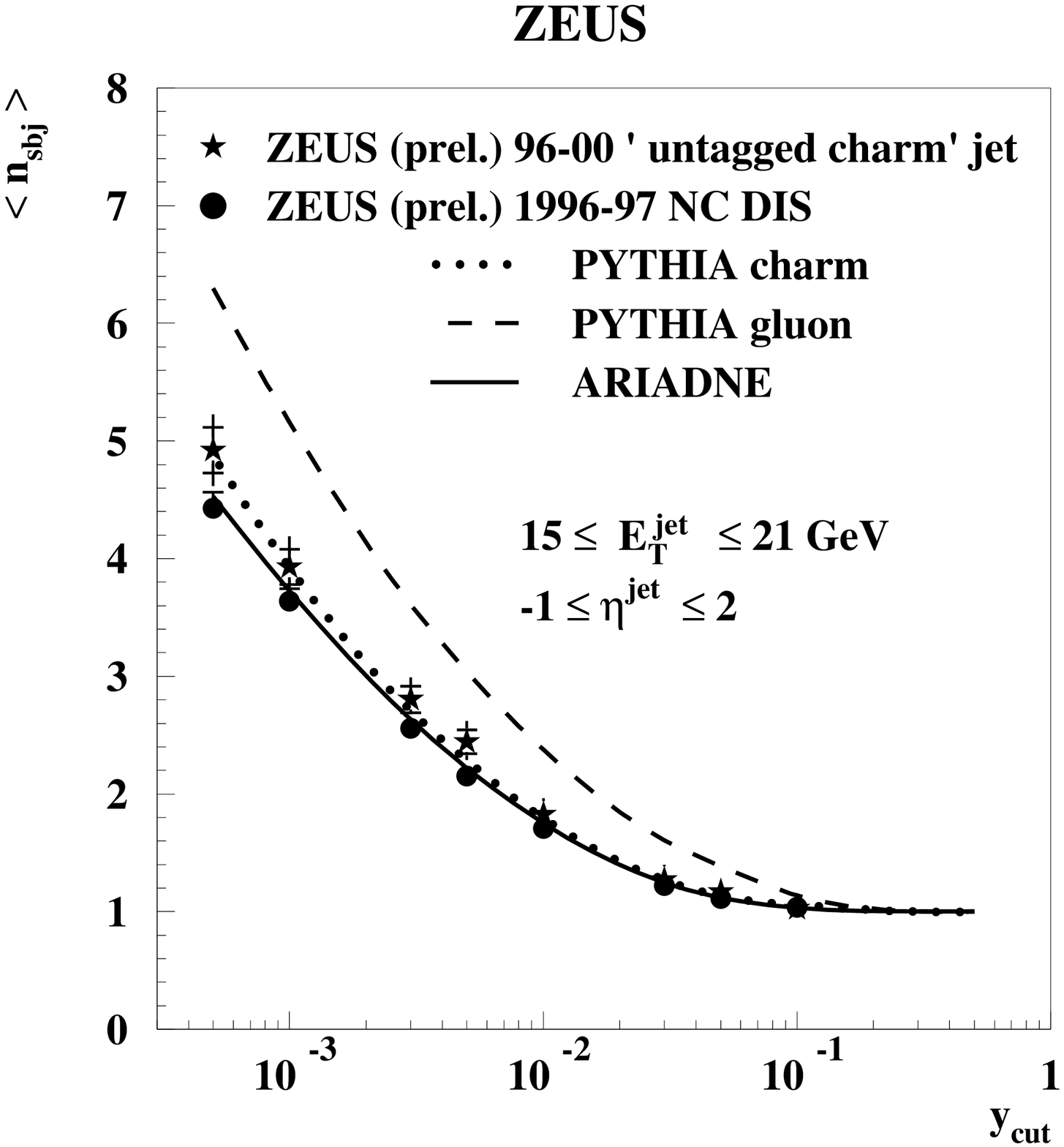,height=6.0cm,clip=}}
    \put(1.,1.){\small (a)}
  \end{picture}
\end{minipage}
\hfill
\begin{minipage}{6cm}
  \begin{picture}(5,6)
    \put(0.,3.5){\psfig{file=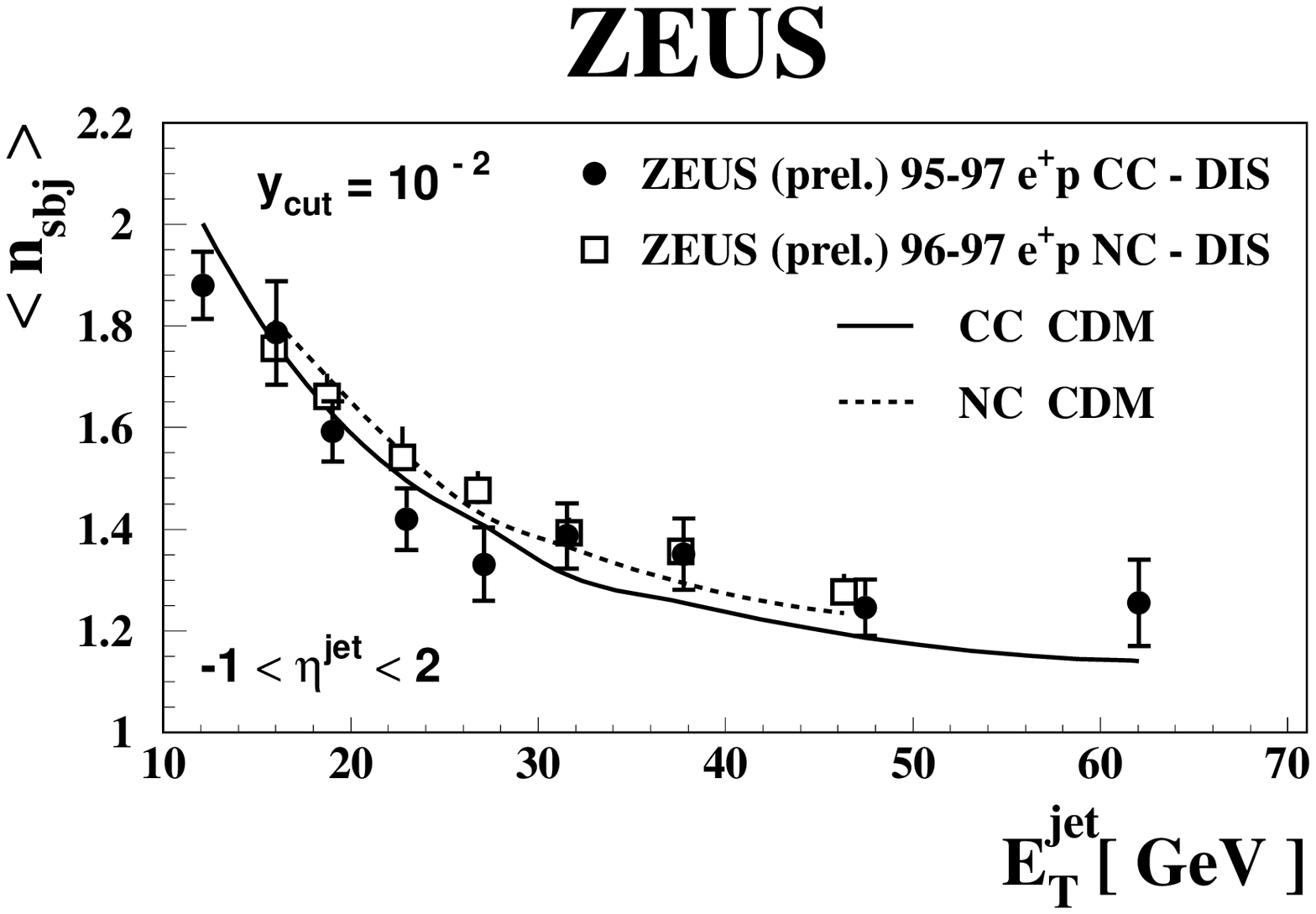,width=6.0cm,clip=}}
    \put(0.,0.){\psfig{file=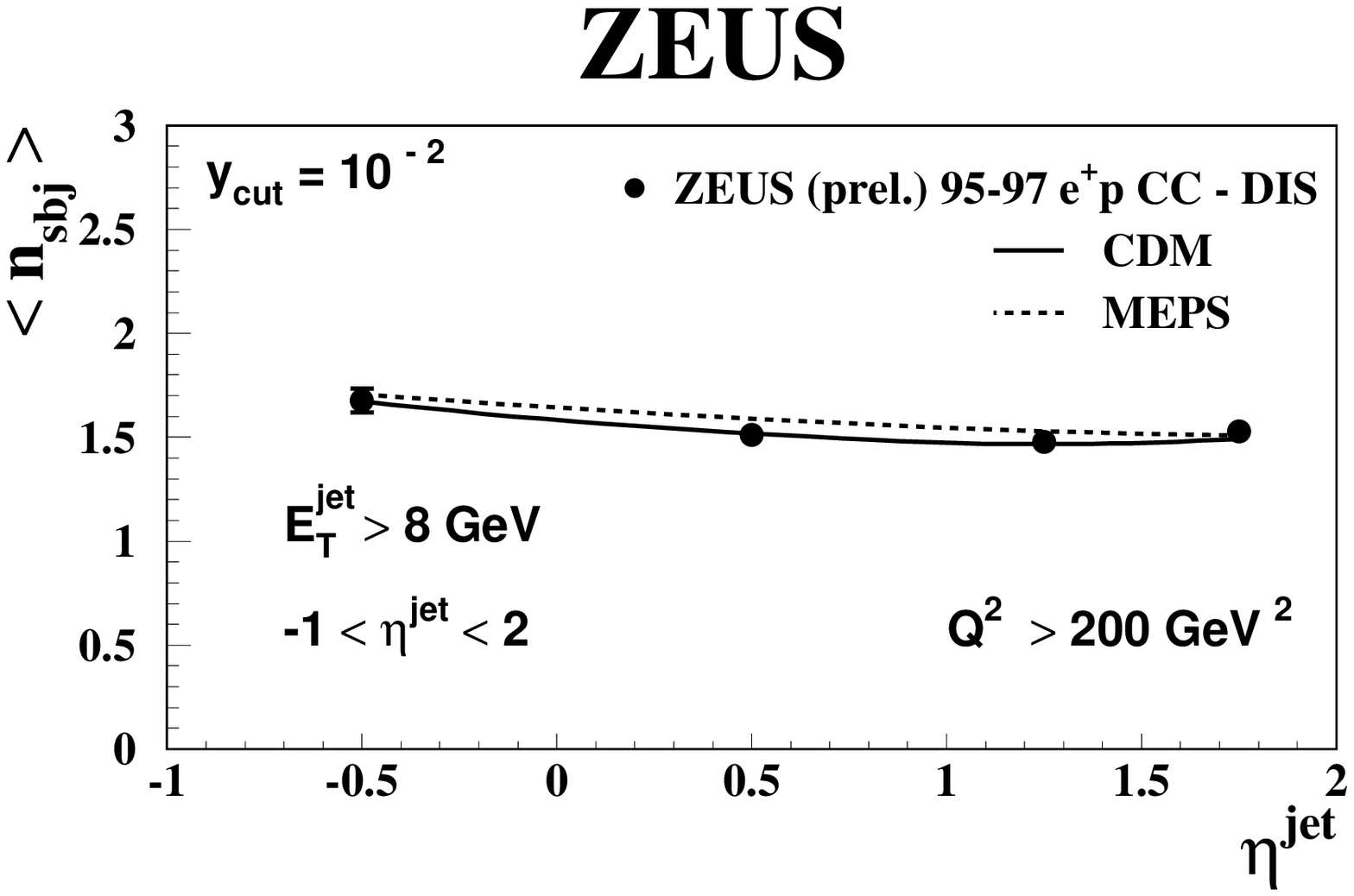,width=6.0cm,clip=}}
    \put(3.,4.5){\small (b)}
    \put(3.,1.){\small (c)}
  \end{picture}
%
\end{minipage}
\vspace*{-0.4cm}
\caption{
(a) Comparison of the subjet multiplicity as a function of $y_{cut}$ corrected to hadron level for the ``untagged-charm'' jet PHP sample and inclusive jet production in NC DIS . (b) Comparison of $\langle n_{sbj}\rangle$ for a fixed $y_{cut}=0.01$ as a function of $E_T^{jet}$ for inclusive jet production in NC and CC DIS. (c) Dependence of $\langle n_{sbj}\rangle$ with $\eta^{jet}$ for a fixed $y_{cut}=0.01$ in CC DIS.
\label{comp}
}
\vspace*{-0.7cm}
\end{figure}

Since the dijet PHP sample consists of a mixture of quark and gluon jets, any measured observable ${\cal{O}}$ of the internal structure can be written as: ${\cal{O}}_{dijet}=f_q \cdot {\cal{O}}_{quark} + f_g \cdot {\cal{O}}_{gluon}$, where $f_q$, $f_g=1-f_q$ are the fractions of quark and gluon jets. The measurements of the substructure of the charm-enriched sample at high transverse energies ($E_T^{jet}>15$ GeV) can be considered as measurements for a pure sample of quark jets (${\cal{O}}_{charm}={\cal{O}}_{quark}$). Taking the fractions $f_q$, $f_g$ from LO Monte Carlo, the substructure of gluon jets can be extracted. The fraction $f_q$ predicted for a dijet sample with $E_T^{jet}>15$ GeV and $-1<\eta^{jet}<2$ by PYTHIA ($f_q=0.66$) and HERWIG~\cite{herwig}($f_q=0.69$) are found to be similar. Fig.~\ref{sub_eta}b shows the extracted gluon jet substructure, which is consistent with the QCD predictions.
\vspace*{-0.4cm}
\section{Jet substructure in deep inelastic scattering}
\vspace*{-0.3cm}
Quark initiated jets are expected to be predominant in charged current (CC) and neutral current (NC) deep inelastic scattering (DIS). In Fig.~\ref{comp}a, $\langle n_{sbj}\rangle$ as a function of $y_{cut}$ in the ``untagged-charm jet'' PHP sample and an inclusive jet NC sample ($Q^2>125$ GeV$^2$) are compared. The measurements are found to be similar and in very good agreement for resolution scales $y_{cut}>0.01$, where the charm mass effects are negligible. The charm-initiated jets are very similar to light quark jets. Fig.~\ref{comp}b shows the evolution of $\langle n_{sbj}\rangle$ with $E_T^{jet}$ for a fixed $y_{cut}=0.01$ in NC and CC processes. The value of $\langle n_{sbj}\rangle$ decreases as $E_T^{jet}$ increases. The agreement between both measurements indicates that the pattern of parton radiation within quark jets is to a large extent independent of the hard scattering process. In Fig.~\ref{comp}c $\langle n_{sbj}\rangle$  for a fixed $y_{cut}=0.01$ as a function of $\eta^{jet}$ in CC interactions ($Q^2>200$ GeV$^2$)  is shown. Both ARIADNE (CDM)~\cite{ariadne} and LEPTO (MEPS)~\cite{lepto} give a good description of the data. The substructure in DIS processes shows no dependence with $\eta^{jet}$. 

\begin{figure}[t]
  \unitlength 1cm
\vspace*{-2.cm}
\begin{minipage}{6.2cm}
  \begin{picture}(5,5)
    \put(0.,0.){\psfig{file=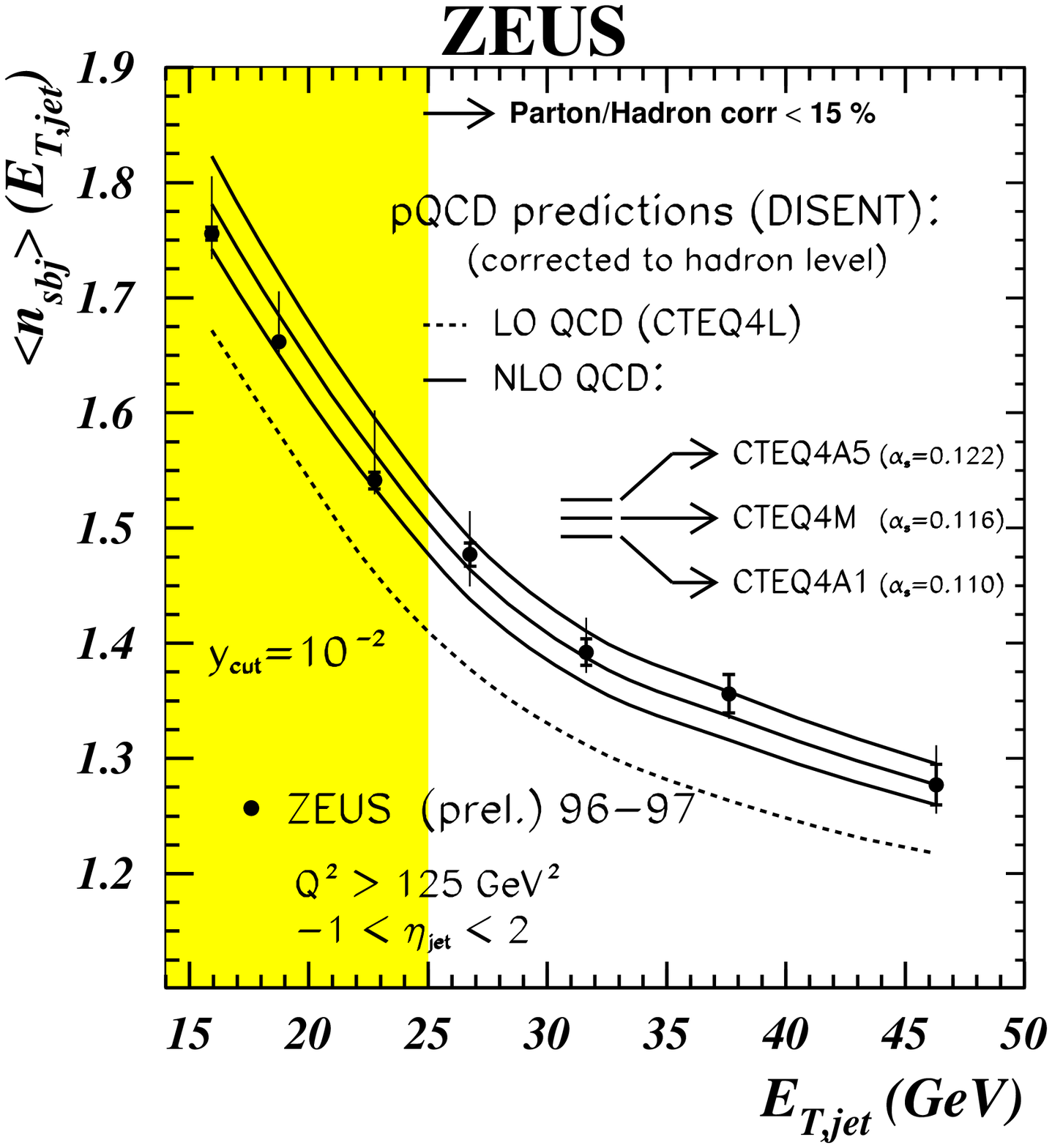,width=6.2cm,clip=}}
    \put(4.5,1.2){(a)}
  \end{picture}
\end{minipage}
\hfill
\begin{minipage}{6cm}
  \begin{picture}(5,5)
    \put(0.,0.){\psfig{file=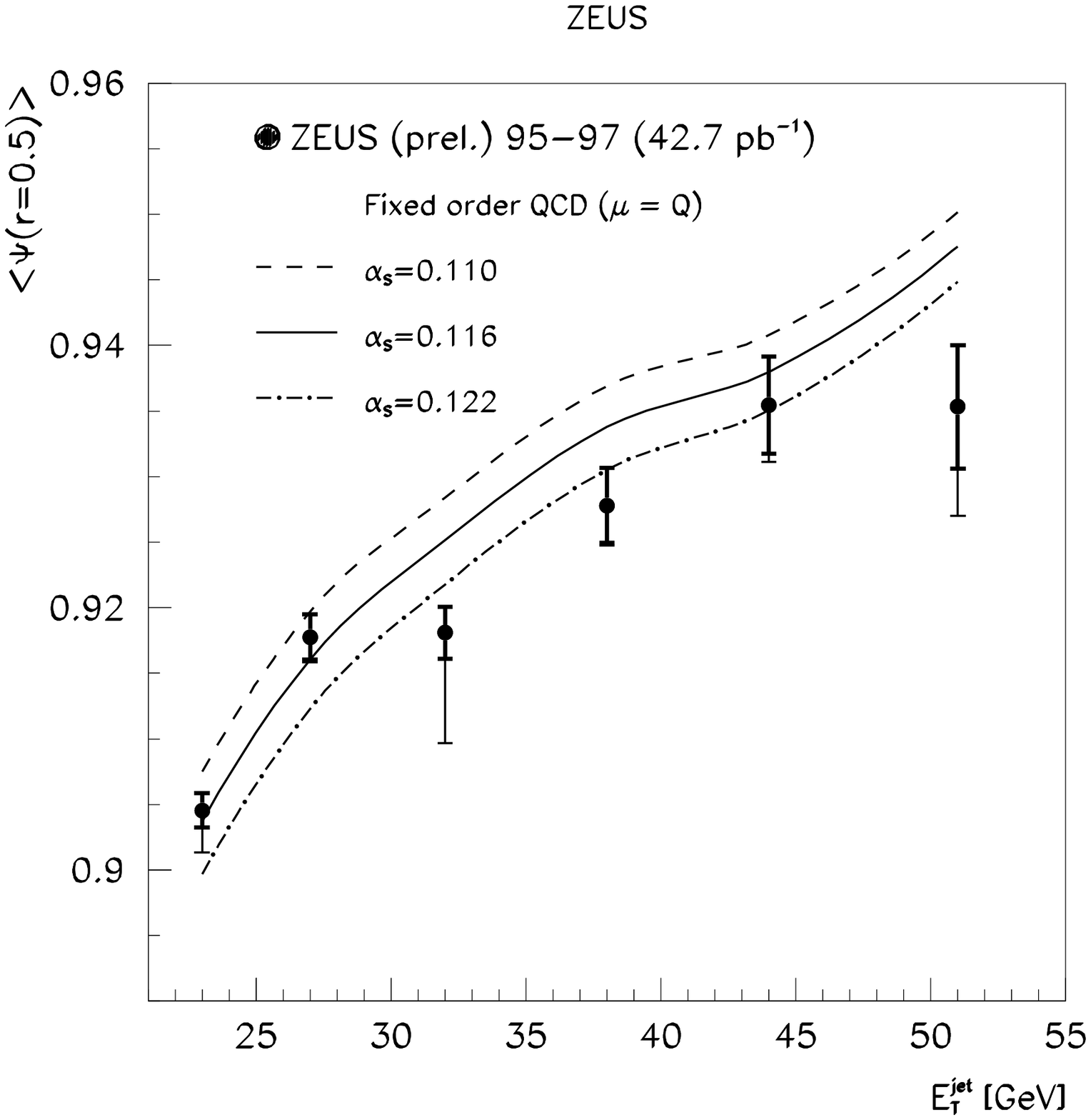,width=6.0cm,clip=}}
    \put(4.5,1.2){(b)}
  \end{picture}
\end{minipage}
\vspace*{-0.4cm}
\caption{
Measured (a) $\langle n_{sbj}\rangle$ and (b) $\langle \Psi(r)\rangle$ for a fixed $y_{cut}=0.01$ as function of $E_T^{jet}$. The data are compared to NLO QCD calculations for different values of $\alpha_S$.
\label{alpha}
\vspace*{-.8cm}
}
\end{figure}

In NC processes, the subjet multiplicity (Fig.~\ref{alpha}a) as a function of  $E_T^{jet}$ for a fixed $y_{cut}=0.01$ and the mean integrated jet shape (Fig.~\ref{alpha}b) for a fixed radius of $r=0.5$  have been compared to NLO QCD  predictions  and $\alpha_S$ has been determined. The extracted value from $\langle n_{sbj}\rangle$ for jets with $E_T^{jet}>25$ GeV is 
$\alpha_S(M_Z) = 0.1185 \pm 0.0016 (stat) ^{+0.0067}_{-0.0048}(exp) ^{+0.0089}_{-0.0071}(th)$.
\vspace*{.1cm}
\newline
The extracted value from $\langle\Psi(r)\rangle$ for jets with $E_T^{jet}>21$ GeV is 
\vspace*{.1cm}
\newline
$\alpha_S(M_Z) = 0.1179 \pm 0.0017 (stat) ^{+0.0054}_{-0.0065}(exp) ^{+0.0094}_{-0.0073}(th)$.
 \vspace*{.1cm}
\newline
Both values are in good agreement with the world average value~\cite{pdg}.

\vspace*{-0.3cm}

\vspace*{-0.5cm}
\begin{thebibliography}{99}
\vspace*{-0.3cm}
\bibitem{sh} S.D. Ellis, Z. Kunszt and D.E. Soper {\it Phys. Rev. Lett.} {\bf 69}, 3615 (1992).
\bibitem{su} S. Catani et al., {\it Nucl. Phys. }B {\bf 377}, 445 (1992) and {\it Nucl. Phys. }B {\bf 383}, 419 (1992); M.H. Seymour, {\it Nucl. Phys. }B {\bf 421}, 545 (1994) and {\it Phys. Rev. Lett.} B {\bf 378}, 279 (1996).
\bibitem{kt} S. Catani et al.,{\it Nucl. Phys. }B {\bf 406},187 (1993); S.D. Ellis and D.E. Soper, {\it Phys. Rev.} D {\bf 48}, 3160 (1993).
\bibitem{pythia} T. Sj\"{o}strand, {\it Comput. Phys. Commun.} {\bf 82}, 74 (1994).

\bibitem{herwig} G. Marchesini et al., {\it Comput. Phys. Commun.} {\bf 67}, 465 (1992).

\bibitem{ariadne} L. L\"{o}nnblad et al., {\it Comput. Phys. Commun.} {\bf 71}, 15 (1992).

\bibitem{lepto} G. Ingelman et al., {\it Comput. Phys. Commun.} {\bf 101}, 108 (1997).
\bibitem{pdg} Particle Data Group, {\it Eur. Phys. J.} {\bf C 15}, 1 (2000).
\end{thebibliography}
\end{document}